\documentstyle[aps,prb,twocolumn,floats,psfig]{revtex}

\draft
\begin{document}

\twocolumn[\hsize\textwidth\columnwidth\hsize\csname@twocolumnfalse%
\endcsname

\title{Absence of effective spins 1/2 induced by nonmagnetic 
impurities in a class of low-dimensional magnets}


\author{B. Normand$^1$ and F. Mila$^2$}

\address{$^1$Theoretische Physik III, Elektronische Korrelationen und 
Magnetismus, Universit\"at Augsburg, D-86135 Augsburg, Germany \\
$^2$Institut de Physique Th\'eorique, Universit\'e de Lausanne,
CH-1015 Lausanne, Switzerland}

\date{\today}
\maketitle

\begin{abstract}

We show that doping a Majumdar--Ghosh chain with non-magnetic impurities 
does {\it not} produce almost free spins 1/2. The difference between this 
system and other spin liquids, such as unfrustrated spin ladders, is 
illustrated in the context of the general dimerized, frustrated spin chain. 
By detailed analysis of the excitation spectra of finite chains with two 
impurities, we investigate the evolution of the screening affecting 
impurity-induced free spins, and of the sign and magnitude of their 
effective interactions. We comment on a possible connection to 
impurity-doping experiments in CuGeO$_3$. 

\end{abstract}
\pacs{PACS numbers: 75.10.Jm, 75.30.Hx, 75.40.Mg, 75.50.Ee}
]

\section{Introduction}

Doping of low-dimensional spin systems by static, nonmagnetic impurities 
has been a topic of increasing interest in recent years, driven by the 
observation that significant information may be obtained concerning 
the nature of magnetic fluctuations. For antiferromagnets, the issue 
was first addressed in the Heisenberg model,\cite{rbhsl} where a small 
induced moment was found around each empty site. A comprehensive 
study of Hubbard antiferromagnets\cite{rudjssvz} suggested that the size 
of the effective moment may vary from this small value in the limit of 
large $U/t$ to a full compensation of the missing spin at small $U/t$. 
Nonmagnetic impurities in high-temperature superconductors 
beyond the antiferromagnetically ordered region at low hole doping were 
found in nuclear magnetic resonance (NMR) measurements to give localized 
moments of $S = 1/2$,\cite{rmacm} while further NMR studies have investigated 
the effects of impurities on magnetic correlations,\cite{rbmambcm} and shown 
that each vacancy causes a strong antiferromagnetic (AF) polarization of its 
immediate vicinity.\cite{rjfhbbscm} A full theoretical discussion can be 
found in Ref.~\onlinecite{rvbs}.

Nonmagnetic impurities in quantum spin liquids give rise to various 
phenomena, among which early attention was focused on the formation 
at low temperature of large-spin clusters with random, effective 
interactions\cite{rhrg,rwfsl,rsf} on a subsystem of free spins isolated 
by doping. For ladder\cite{rsf} and dimerized\cite{rfts} geometries 
these interactions are coherent and quasi-long-ranged, in accord with 
the weak antiferromagnetism observed in the low-temperature phases obtained 
on Zn-doping of the quasi-1d ladder material $\rm SrCu_2O_3$,\cite{raftnt} 
and of the dimerized chain CuGeO$_3$.\cite{rskumhhs} These ideas have been 
extended to discuss true magnetic order on the impurity spin subsystem in 
higher dimensions for coupled, dimerized chains\cite{rdhrap} and for the 
1/5-depleted square lattice,\cite{rwnsh} both of which are bipartite. More 
generally, Martins {\it et al.}\cite{rmlrd} argued for local enhancement 
of AF correlations around impurity sites in a range of low-dimensional $S 
= 1/2$ systems, and later discussed the evidence for nonmagnetic impurities 
liberating effectively free spins,\cite{rlmgmdhlr} a notion which now has
 broad acceptance for all spin liquids. 

\begin{figure}[ht]
\vspace{0.3cm}
\centerline{\psfig{figure=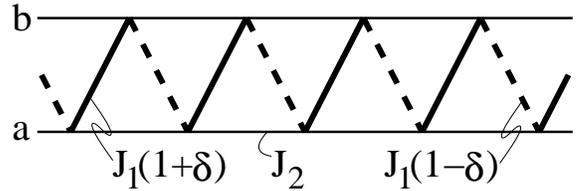,width=7.5cm,angle=0}}
\caption{ The dimerized, frustrated spin chain.} 
\end{figure}

However, very few of these studies have considered systems with magnetic 
frustration. In this paper, we show that in frustrated systems missing 
moments do not necessarily produce nearly-free spins 1/2, even when the 
spin gap is large. The analysis is based on the Heisenberg model for a 
chain with dimerized nearest-neighbor and frustrating next-nearest-neighbor 
interactions (Fig.~1), described by the Hamiltonian
\begin{eqnarray}
H & = & J_1 \sum_i \left[ (1 + \delta){\bf S}_i^{\rm a} {\bf \cdot S}_i^{\rm 
b} + (1 - \delta) {\bf S}_i^{\rm b} {\bf \cdot S}_{i+1}^{\rm a} \right] 
\label{esh} \nonumber \\ & & \;\;\;\; + J_2 \sum_{i, {\rm i;m = a,b}} {\bf 
S}_i^{\rm m} {\bf \cdot S}_{i+1}^{\rm m} .
\end{eqnarray}
We perform zero-temperature Lanczos diagonalization of finite $S = 1/2$ 
chains containing two $S = 0$ impurities, and from inspection of the 
low-energy excitation spectrum deduce the nature of the impurity-induced 
ground state. All of the results to follow were obtained on periodic 
chains of 20 sites, meaning with 18 spins and 2 impurities, but we have 
verified with longer chains that the results on which we will base our 
conclusions are dominated by the shorter inter-impurity separation, and 
remain essentially independent of the chain length. 


\begin{figure}[ht]
\centerline{\psfig{figure=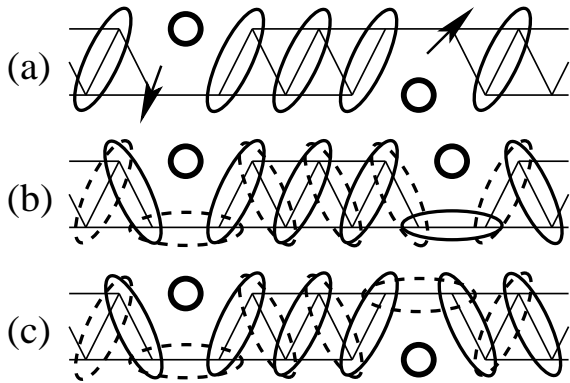,width=7.5cm,angle=0}}
\vspace{0.3cm}
\caption{Schematic representation of nearest-neighbor dimer singlet coverings 
in the ladder (a) and in the frustrated chain (b,c) with two impurities, 
represented by the open circles.}
\end{figure}

\section{Spin ladder}

We begin by illustrating the nature of our analysis for the well-known 
case of the spin ladder, which is the $\delta = 1$ limit of Eq.~(\ref{esh}). 
As shown in Fig.~2(a), each impurity is expected to leave an effectively 
free spin on the depleted rung. By ``effectively free spin'' we understand 
an $S = 1/2$ degree of freedom whose interactions with any other moments in 
the system occur on an energy scale very much smaller than the superexchange 
interactions in Eq.~(\ref{esh}). In a ladder the interactions between the 
free spins are exponentially weak because of the spin gap and consequent 
short correlation length, and have an effective AF or ferromagnetic (FM) 
nature according to the Marshall sign rule.\cite{rm,rsf} In Fig.~3(a) are 
shown the lowest four energy levels for ladders with all chain-to-rung 
coupling ratios from 0 to 1, for a configuration with the first impurity 
at (1,b) (first rung, b-chain; see Fig.~1) and the second at (4,b). Because 
these are on opposite sublattices, the lowest level is a singlet, while the 
first excited state is a triplet, and there is always a wide separation 
of this lowest manifold from other levels. This situation is mirrored for 
the impurity configuration ((1,b)(4,a)) [Fig.~3(b)], but with singlet and 
triplet levels interchanged. We interpret these results as corresponding 
to two free spins isolated by the impurities, and communicating with 
effective AF (a) or FM (b) interactions given by the singlet-triplet 
separation in the lowest manifold. 

\begin{figure}[ht]
\centerline{\psfig{figure=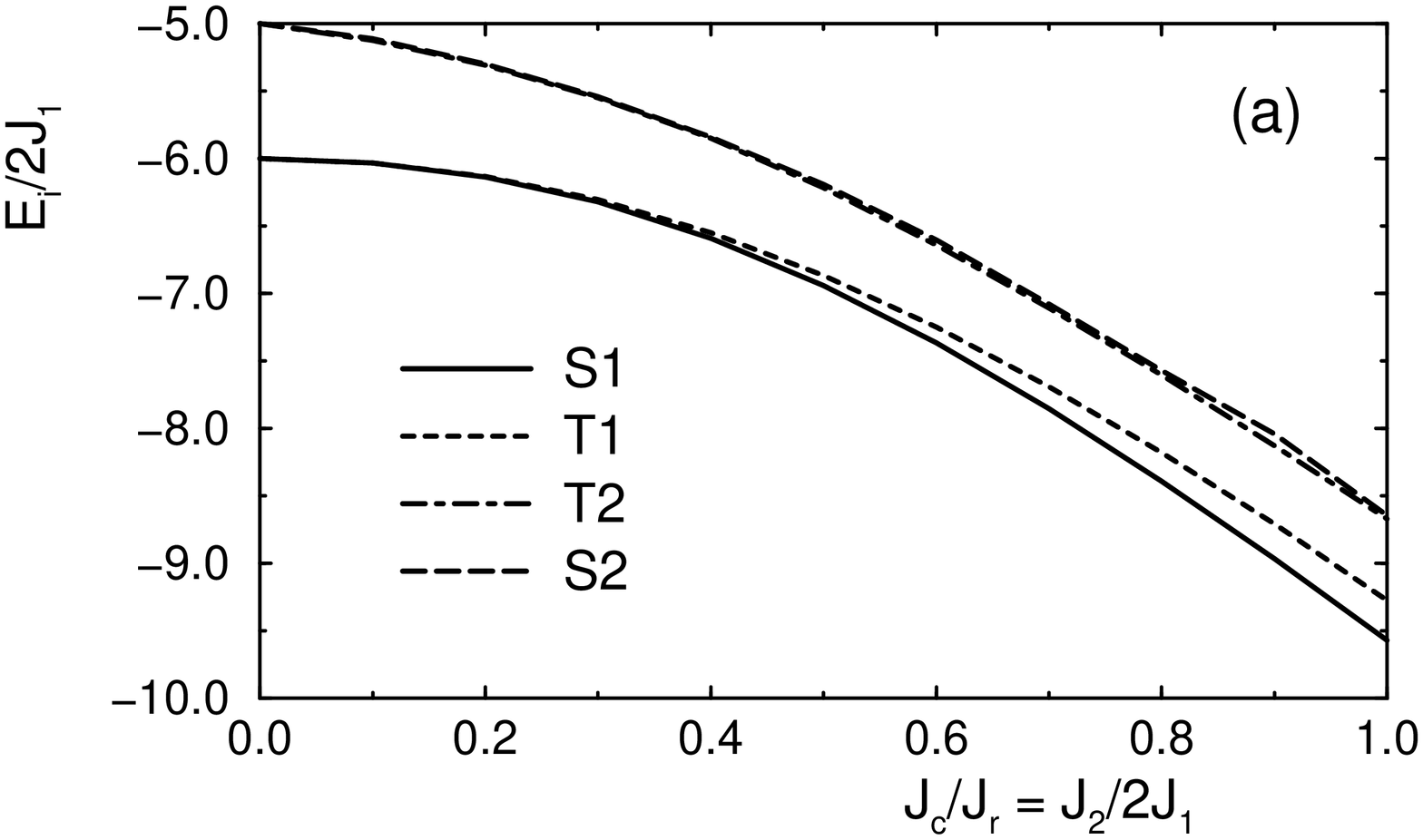,width=8.5cm,angle=270}}
\centerline{\psfig{figure=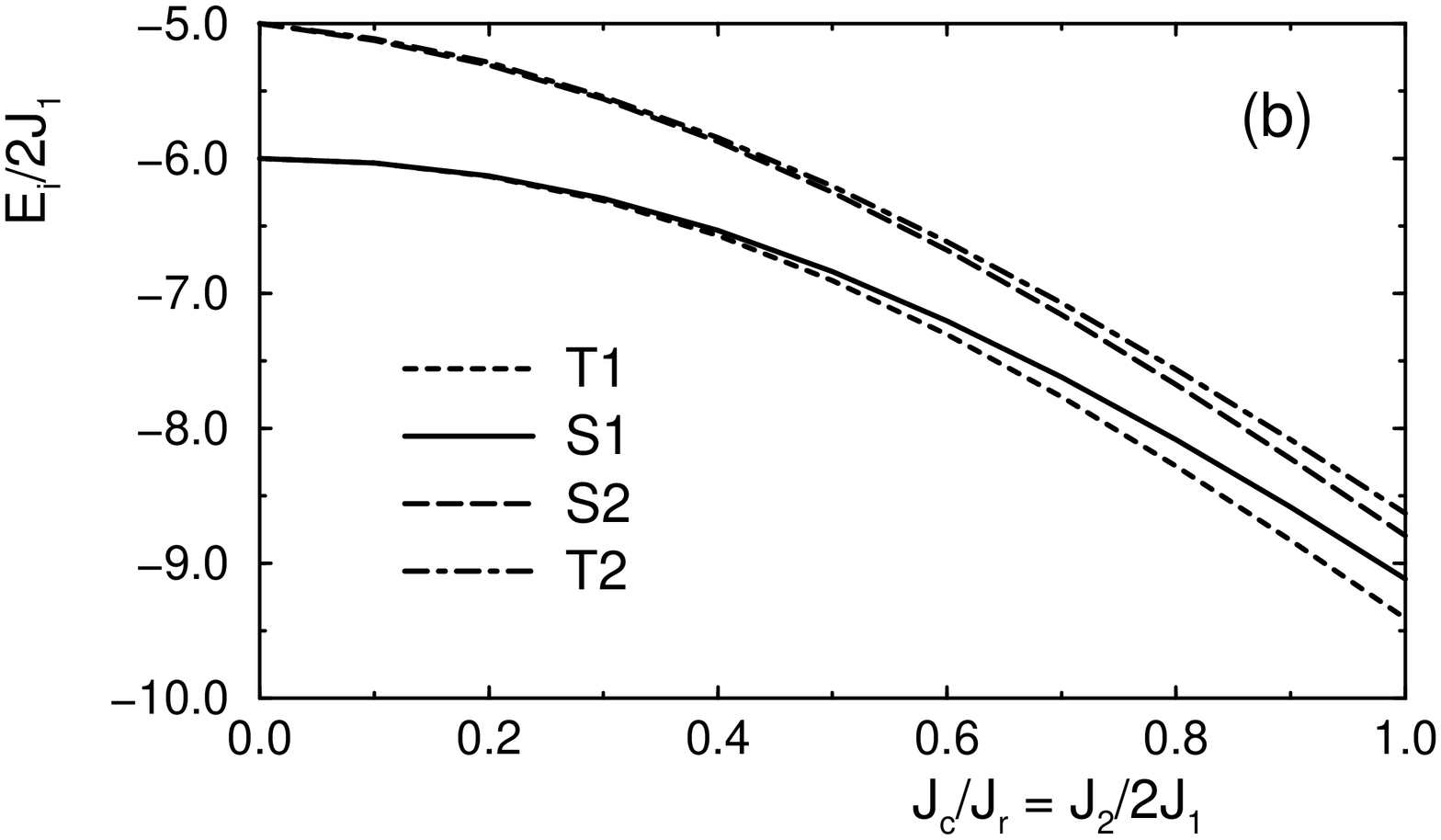,width=8.5cm,angle=270}}
\vspace{0.0cm}
\caption{Lowest four energy levels $E_i$ as a function of the chain to rung 
coupling ratio, $J_{\rm c}/J_{\rm r}$, for the spin ladder with 
impurities on the first and fourth rungs. In (a) the impurities occupy the 
same chain and in (b) they are located on opposite chains. The labels denote 
first and second singlet and triplet states. For the ladder defined by 
Eq.~(\protect{\ref{esh}}) with $\delta = 1$, $J_{\rm c}/J_{\rm r} = J_2 
/ 2 J_1$.}
\end{figure}

\begin{figure}[ht]
\centerline{\psfig{figure=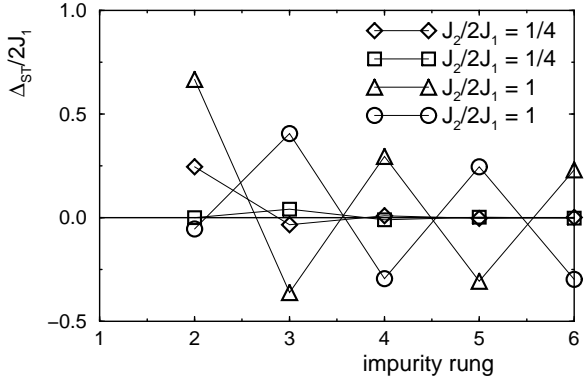,width=8.5cm,angle=270}}
\caption{Singlet-triplet energy gap $\Delta_{\rm ST}$ for ladders with two 
impurities. The first impurity is located at (1,b), and the second on the 
rung indicated on the $x$-axis, on the same (diamonds, triangles) or the  
opposite (squares, circles) chain.}
\end{figure}

In Fig.~4 we show the singlet-triplet gap $\Delta_{\rm ST}$ as a function of 
impurity separation for ladders with $J_2/2 J_1 = 0.25$ and 1. The alternation 
of the sign of $\Delta_{\rm ST}$ is clear, as is the decay in its magnitude 
with separation. For $J_2/2 J_1 = 0.25$ this decay is very rapid, in keeping 
with the expectation of a short coherence length on an undoped ladder with a 
large singlet-triplet gap. We note that the gap values with impurities on 
adjacent rungs correspond approximately to two sites with effective coupling 
$J_2$ when they occupy the same chain, but become very small when they are 
on opposite chains as in this case the ladder is cut. For the isotropic 
ladder, $J_2/2 J_1 = 1$, the decay is no longer clearly exponential, and 
implies that despite the spin gap in this system the impurity-induced states 
have a significant spatial extent which is approaching the system size. Our 
studies of the energy spectra provide a clear and complete confirmation of 
the results of Sigrist and Furusaki\cite{rsf} for the sign and magnitude 
of effective interactions between free spins in the ladder. 

\section{Frustrated chain}

We turn now to the frustrated chain, and begin with the case of no 
dimerization ($\delta = 0$). Much is known about this system, including the 
presence of a quantum critical point separating gapless and gapped phases 
at $J_2/J_1 = 0.2412$, and the existence of an exact wavefunction at the 
Majumdar-Ghosh (MG) point $J_2/J_1 = 1/2$. The gapped phase arises due to 
spontaneous dimerization, and possesses a double degeneracy on translation.
It is immediately apparent from Figs.~2(b,c) that the degenerate dimer
patterns can simply adapt to accommodate a single impurity, while two 
impurities may fall in one of two ways. If their separation is even, the 
odd number of sites between them has no ideal covering (one dimer must 
always fall on a $J_2$ bond), and two nearly-degenerate singlet states 
are expected [represented by the solid and dashed dimers in Fig.~2(b)]. 
If the separation is odd, the even number of sites between the impurities 
can be ideally covered by the solid dimers in Fig.~2(c), whereas the second 
singlet state represented by the dashed dimers is driven to higher energy. 

\begin{figure}[ht]
\centerline{\psfig{figure=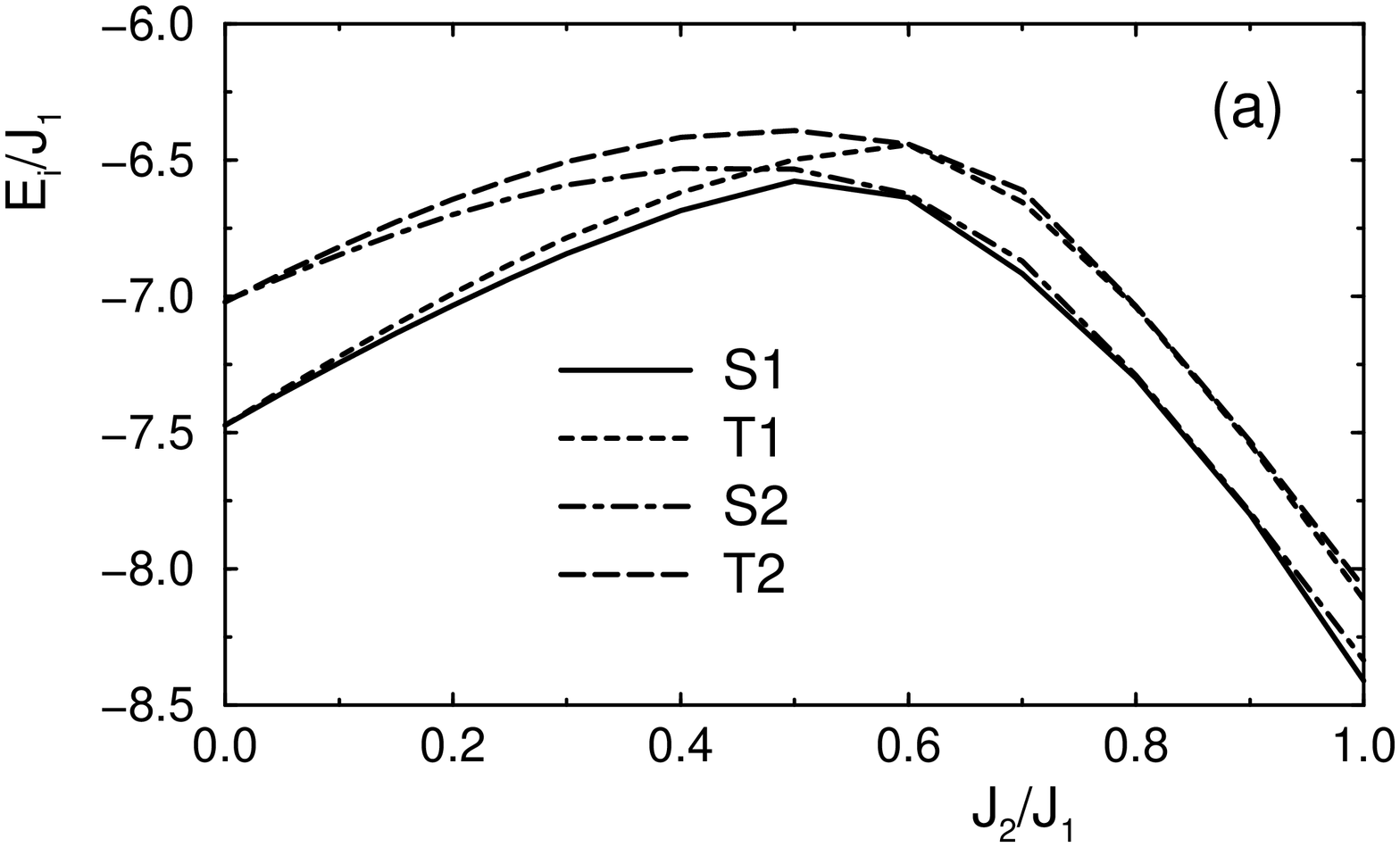,width=8.5cm,angle=270}}
\centerline{\psfig{figure=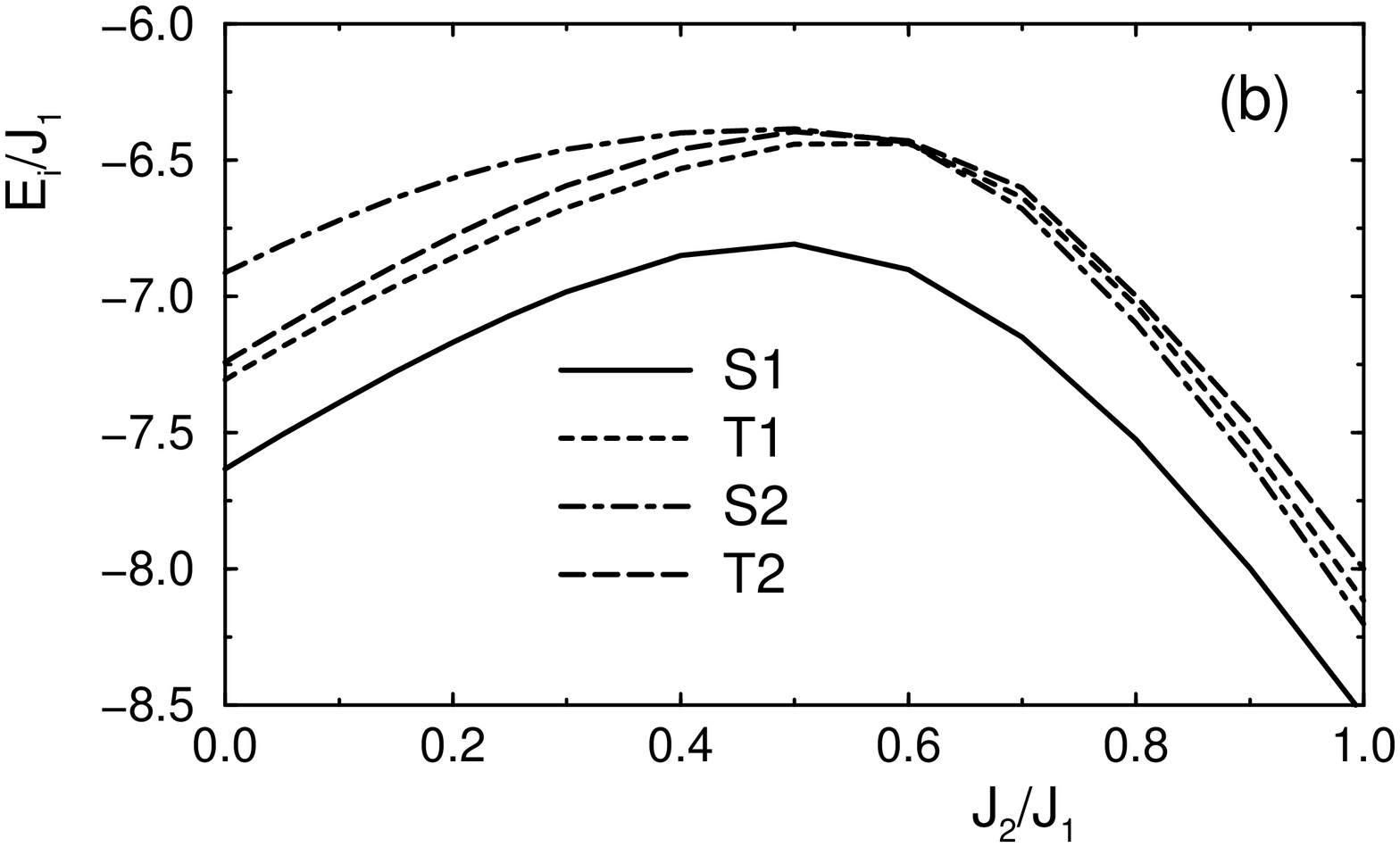,width=8.5cm,angle=270}}
\caption{Lowest four energy levels $E_i$ as a function of frustration 
$J_2/J_1$ for the undimerized chain with two impurities whose separation 
is even (a) or odd (b). The labels denote first and second singlet and 
triplet states.}
\end{figure}

\begin{figure}[ht]
\centerline{\psfig{figure=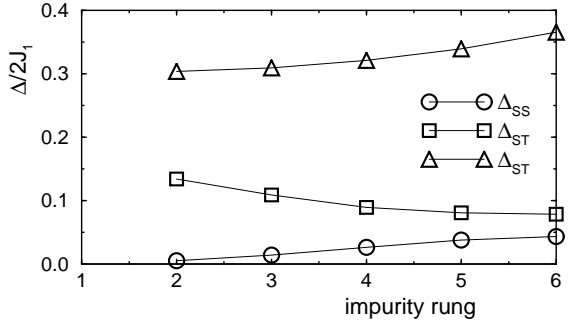,width=8.5cm,angle=270}}
\caption{Singlet-triplet energy gap $\Delta_{\rm ST}$ for MG chains ($J_2 
/ J_1 = 0.5$) with two impurities. The first impurity is located at (1,b), 
and the second on the rung indicated on the $x$-axis, separated by an even 
(squares) or an odd (triangles) number of sites. Also shown (circles) is 
the energy $\Delta_{\rm SS}$ of the first excited singlet state, which is 
nearly degenerate with the ground state for even impurity separation. }
\end{figure}

Fig.~5(a) shows the lowest levels in the energy spectrum for the case 
of impurities located on sites (1,b) and (6,b), a situation analogous 
to Fig.~2(b). For $J_2/J_1 \ge 1/2$ we do indeed find a pair of low-lying 
singlets as expected from the above argument, and in fact this manifold 
lies well below the closest triplet states for all values of the 
frustration away from $J_2/J_1 = 1/2$. However, below the MG point there 
is an abrupt level crossing such that the lowest manifold consists of a 
singlet and a triplet. These become degenerate at $J_2 = 0$, which 
corresponds to the limit of the doped Heisenberg chain: in this case the 
chain is simply decoupled into two odd-length segments, each an effective 
$S = 1/2$ degree of freedom, while if the impurity separation is odd (below) 
the two even-length segments are $S = 0$. For a long chain randomly doped 
with a finite concentration of impurities, one observes on average half a 
free spin per impurity.\cite{rmmb} The results for frustration values of 
$0 < J_2/J_1 < 0.45$ imply that the chain segments lose the resonant 
character which permits two low-lying singlets, resulting in a pair of 
free spins with a single effective coupling which decreases linearly with 
$J_2$. 

Fig.~5(b) shows the spectrum for the case of impurities with odd spacing, 
where one singlet state lies well below all of the other levels over the 
entire range of frustration, also as expected from the above considerations. 
For appreciable frustration this is the favored (solid) singlet state of 
Fig.~2(c), while for weak $J_2$ it has more the character of two separated, 
short chains of even length. In no circumstances do the impurities introduce 
free spins. Thus for a random distribution of impurities, strongly frustrated 
chains will not show free spin degrees of freedom, but would exhibit a 
crossover to half a free spin per impurity as frustration is reduced. 

In the light of our opening statement concerning size effects in our 
calculations, it is clear that impurity doping destroys the quantum 
critical point at $J_2/J_1 = 0.2412$, which might have been expected 
to mark the boundary between these two types of behavior, and replaces 
it with a crossover. At very low impurity concentrations, this critical 
value should still mark the onset of the crossover as a function of 
$J_2/J_1$. By contrast, for the high doping concentrations represented 
by our results, the effect of $J_2$ in localizing a spin of 1/2 at either 
end of a chain segment cannot compete with the finite-size gaps in the 
spectra of short segments, the regime of half a free spin per impurity 
is moved to larger $J_2 \sim 0.5$, and the crossover broadened. 

Before leaving the undimerized chain, we present briefly the effects of 
altering the impurity spacing in the MG chain. Fig.~6 displays two 
distinctive features in sharp contrast to the case of the spin ladder 
(Fig.~4). The first is that the alternation of sign in the ladder case 
is no longer present, being replaced by the alternation in character
of the ground state (one favored singlet, or two similarly disadvantaged 
singlets) with odd or even impurity spacing. The second is that the energy 
level separations have no appreciable dependence on impurity spacing 
beyond this alternation, demonstrating clearly the local dimer nature of 
the states involved. We note in passing that while nonmagnetic impurities 
in a spin ladder destroy its spin-liquid nature by creation of states 
within the spin gap associated with the free spins, the dimer reorientation 
in the MG chain effectively acts to preserve the spin-liquid character in 
the sense of retaining a significant singlet-triplet gap. 

\section{Dimerized, frustrated chain}

We turn now to the general case of the dimerized, frustrated chain, with 
which one would wish to explore the change in behavior from the MG regime 
to the ladder regime. We investigate this change by varying the dimerization 
parameter $\delta$ from 0 to 1 at a fixed value of $J_2$. We will illustrate 
our results with the case $J_2 = 0.5$, which connects the MG chain to a 
ladder of chain-to-rung coupling 1/4, and have verified from studies 
elsewhere in ($J_2,\delta$) parameter space that the behavior shown is quite 
generic. Fig.~7 gives the low-energy spectra for a selection of impurity 
configurations, presented in the form of the energy level separations 
$\Delta_{\rm ST}$ and $\Delta_{\rm SS}$ between the lowest singlet state, 
which gives the zero of energy, and the first triplet and second singlet, 
respectively. The second triplet level, which is not shown in Fig.~7, lies 
close to the second singlet only at $\delta = 0$ (Fig.~5). 

In Figs.~7(a) and (b) the gaps are shown for impurities on the first 
and fourth rungs, and in Fig.~7(c) for the first and fifth. At small 
dimerization $\Delta_{\rm ST}$ is large, and for impurities on the same 
chain the first excited singlet lies inside this gap. However, this 
level is driven to much higher energies by even rather small dimerization, 
and for most of the parameter range the low-energy manifold contains only 
one singlet and one triplet. The sharp increase in $\Delta_{\rm SS}$ is 
readily understood from the fact that dimerization of the $J_1$ bonds 
limits the flexibility of the system to place nearest-neighbor dimers on 
any link, by penalizing those sections where the dimer pattern preferred 
by the impurity positions lies on the weaker $J_1 (1 - \delta)$ bonds. 

\begin{figure}[t!]
\centerline{\psfig{figure=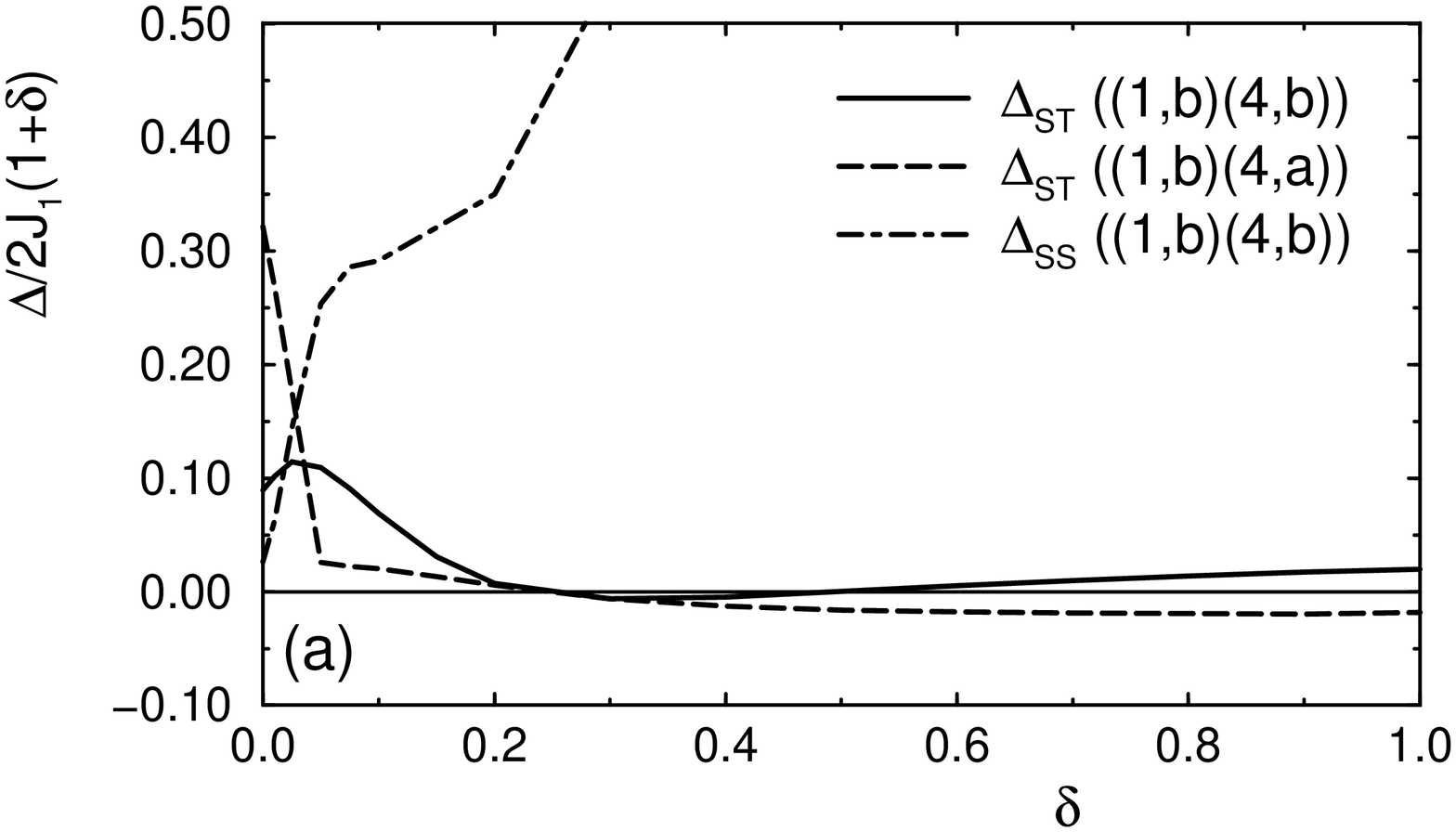,width=8.5cm,angle=270}}
\centerline{\psfig{figure=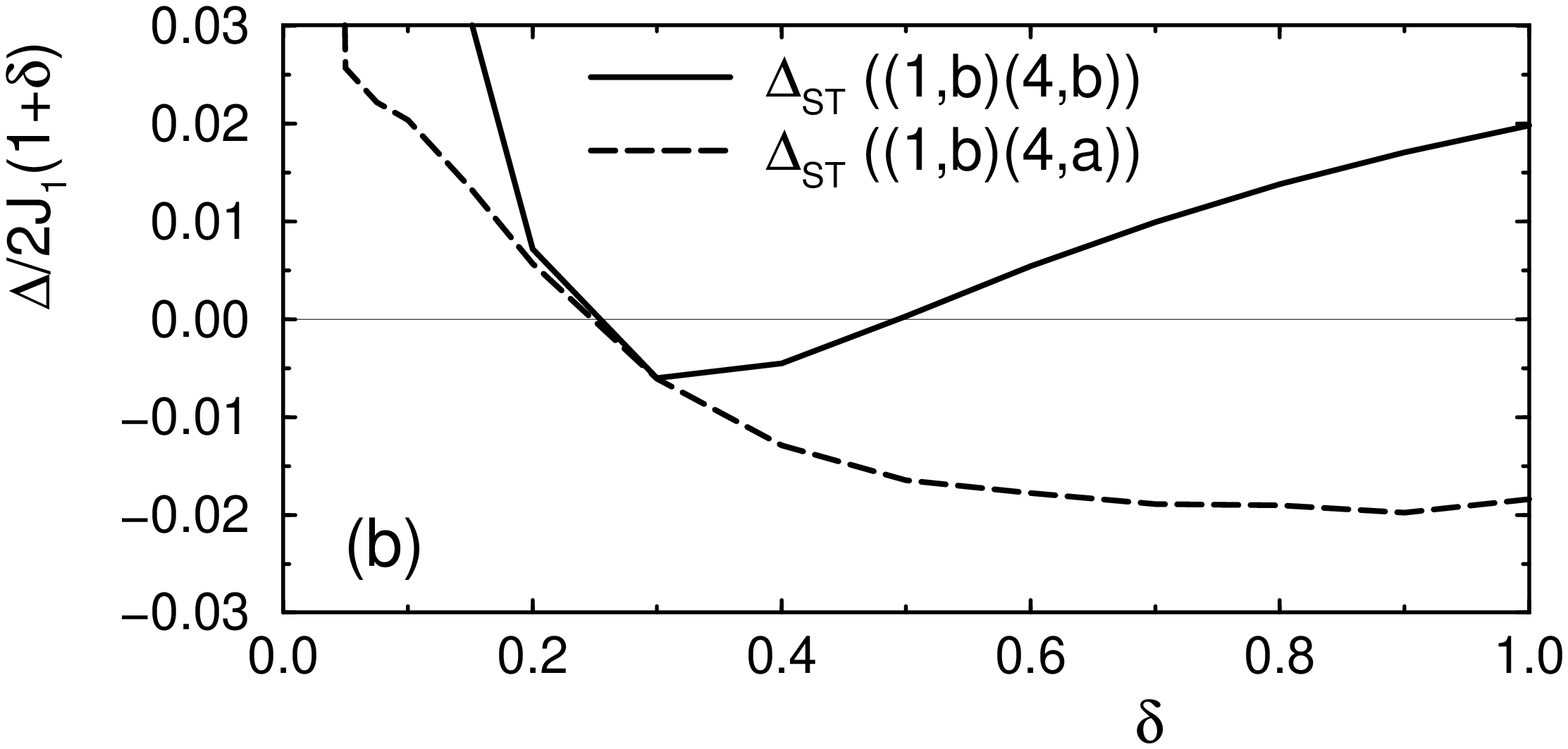,width=8.5cm,angle=270}}
\centerline{\psfig{figure=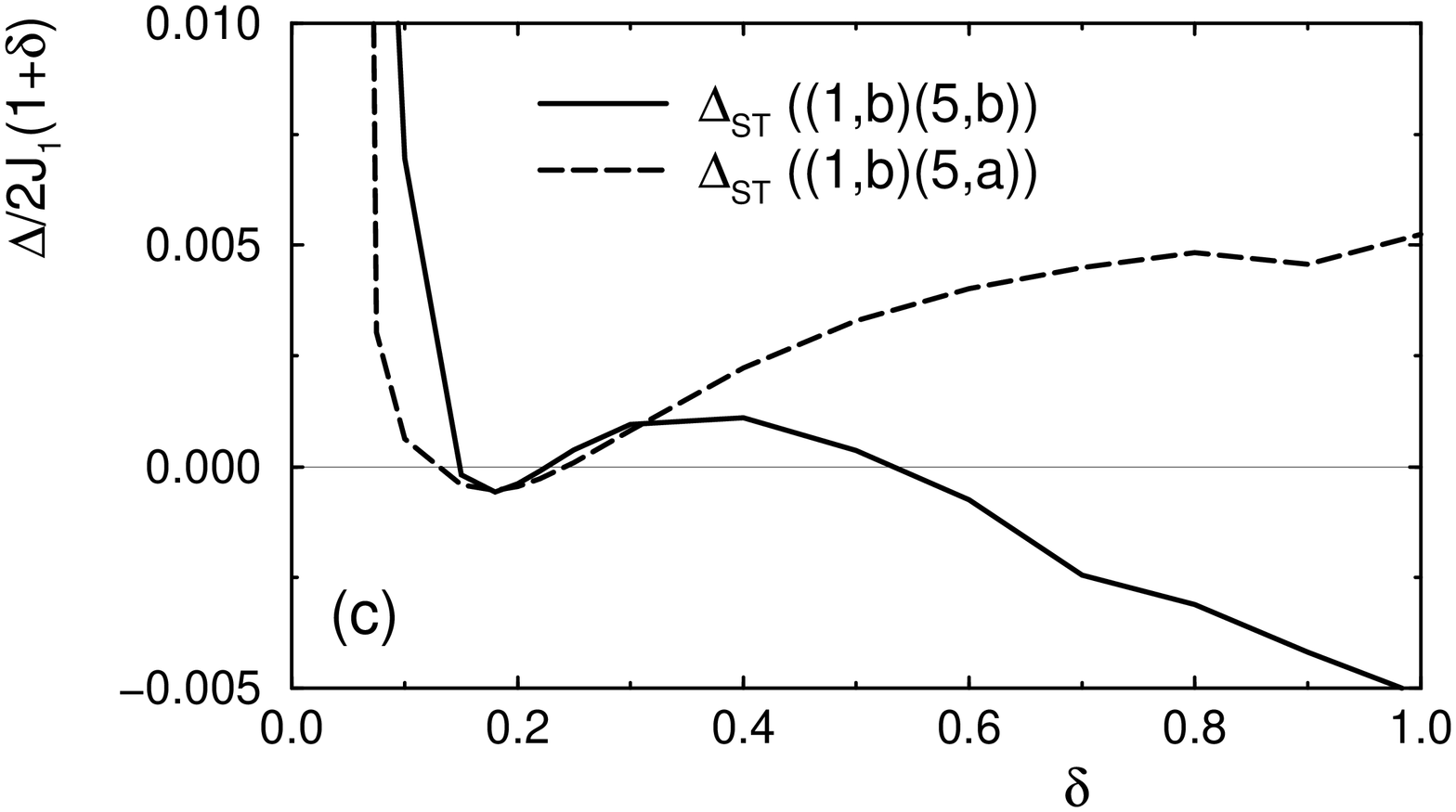,width=8.5cm,angle=270}}
\caption{Singlet-triplet energy-level separation, $\Delta_{\rm ST}$, as a 
function of dimerization $\delta$ for impurities on the first and fourth 
dimers (a,b) and first and fifth (c), at fixed $J_2 = 0.5J_1$. Also shown in 
(a) is the singlet-singlet separation, $\Delta_{\rm SS}$, for the case of 
impurites on the same $J_2$ chain. (b) shows the same data as (a), but 
magnifies the low-energy sector. }
\end{figure}

It is immediately apparent from Fig.~7 that the evolution of the low-lying 
states with dimerization is not monotonic, and as in the ladder limit is 
not governed in magnitude or sign by a single rule. That the ladder- and 
MG-limit ground states do not evolve directly into eachother is no surprise, 
as a significant rearrangement of local dimers is necessary (Fig.~2). 
Similarly, the minimum in the singlet-triplet energy separation at $0.2 
< \delta < 0.3$ arises because reducing $\delta$ from the ladder limit 
increases the frustration between rung singlets and reduces the effective 
interaction until the MG limit is reached at small $\delta$. 

To investigate the nature of the intermediate dimerization regime in 
Fig.~7, we have calculated the expectation values ${\cal S}_m = \langle 
{\bf S}_i {\bf \cdot S}_{i+m} \rangle$ for local dimer formation on nearest- 
($m = 1$) and next-nearest-neighbor ($m = 2$) sites in the chain. In a 
perfect singlet ${\cal S}_1 = -3/4$, and we find that dimer formation on 
the strong bonds $J_1 (1 + \delta)$ is dominant, meaning ${\cal S}_1 \sim 
-0.7$, for all values of the dimerization $\delta > 0.1$. 
The crossover from a MG pattern of flexible dimer formation occurs in 
the range $0.05 < \delta < 0.1$. This is fully consistent with the energy 
spectra (Fig.~7), where the existence of a low-energy manifold containing 
only one singlet and one triplet is established at and above $\delta = 0.1$. 

\begin{figure}[ht]
\centerline{\psfig{figure=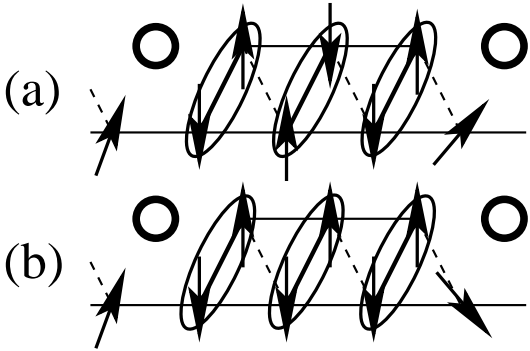,width=7.0cm,angle=0}}
\vspace{0.3cm}
\caption{ Representation of dimer polarization for the dimerized, 
frustrated chain. (a) Ladder limit, inter-dimer coupling determined by 
$J_2$ bonds: ${\cal S}_1 > 0$ between dimers, ${\cal S}_2 < 0$. (b) MG 
chain limit, inter-dimer coupling determined by $J_1 (1 - \delta)$ bonds: 
${\cal S}_1 < 0$ between dimers, ${\cal S}_2 > 0$. }
\end{figure}

A more detailed understanding of the effective interactions can be obtained 
from ${\cal S}_1$ on the bonds between the strong dimers, and from ${\cal 
S}_2$. In a ladder, ${\cal S}_2 < 0$ because of the AF chain bonds, and 
${\cal S}_1 > 0$ as a consequence, although both quantities are small 
[${\cal O}(10^{-2})$]. In a MG chain with two impurities, we find that the 
inter-dimer coupling is controlled by $J_1 (1 - \delta)$ rather than by the 
two bonds $J_2$, and ${\cal S}_1 < 0$ while ${\cal S}_2 > 0$. These two 
situations are represented in Fig.~8. On the quantitative level, even for 
$\delta \sim 0.1$, the inter-dimer $|{\cal S}_1|$ and $|{\cal S}_2|$ are 
of order 0.1, as would be expected from the magnitude of the singlet-triplet 
energy separation (effective inter-impurity interaction). 

Returning to Figs.~7(a) and (b), in the ladder limit $\delta \rightarrow 
1$ the ground state is a singlet (triplet) for impurities on the same 
(opposite) chains. This situation is maintained as $\delta$ is lowered, 
until for the configuration ((1,b)(4,b)) the triplet crosses the singlet 
at $\delta = 0.5$ to become the ground state. This result has a ready 
interpretation in terms of the ``free-spin'' physics which governs the 
behavior of the doped ladder: when the impurities occupy the same chain 
the number of spins between them is odd, and one is a free spin in the 
ladder limit. This spin is coupled to one site of the neighboring dimer 
in the shorter inter-impurity chain segment by $J_2$, and to the other by 
$J_1 (1 - \delta)$ (Fig.~9), so when the latter becomes stronger than the 
former the sign of the effective interaction changes. Note that the other 
free spin does not have a $J_1 (1 - \delta)$ bond to the inter-impurity 
segment, and so the sign of its coupling is unaltered. When the impurities 
are on opposite chains, neither or both, but not only one, can have a $J_1 
(1 - \delta)$ link to the short segment, so that no sign-change can occur. 
The triplet ground states for dimerizations $0.25 < \delta < 0.5$ thus have 
a consistent explanation in terms of free spins introduced by each impurity. 

However, from the previous discussion of dimerization expectation values 
${\cal S}_m$, the singlet regimes directly below $\delta = 0.25$ for impurity 
configurations ((1,b)(4,m)) should not be simply those of the MG limit, and 
this crossover should not mark the onset of the regime governed by 
degeneracy-induced screening of the spins explored in Fig.~5. The computed 
dimerization patterns show that for short inter-impurity segments, while 
dimer formation on the $J_1 ( 1 + \delta )$ bonds remains strong, the 
region around $\delta = 0.25$ does mark the crossover between the two types 
of dimer polarization. For a ((1,b)(4,m)) impurity configuration, where 
there are two dimers in the inter-impurity segment, dimer repolarization 
from a MG ``head-to-tail'' pattern [Fig.~8(b)] to a ladder-like 
``head-to-head'' situation [Fig.~8(a)] accounts directly for the change 
in the spin sector of the ground state seen at $\delta = 0.25$ in Fig.~7(b). 

\begin{figure}[ht]
\centerline{\psfig{figure=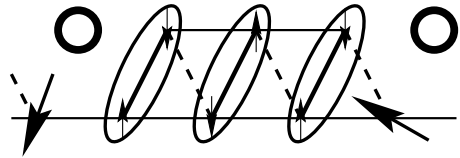,width=6.0cm,angle=0}}
\vspace{0.3cm}
\caption{ Reversal of the effective interaction sign for odd-length 
inter-impurity segments at $\delta = 0.5$: the ``free spin'' on the right 
has competing couplings $J_1 ( 1 - \delta )$ and $J_2$ to the neighboring 
singlet, which is (exponentially) weakly polarized by the other free spin. } 
\end{figure}

In Fig.~7(c), the behavior in the MG regime is identical to Fig.~7(a) (see 
Fig.~6), and so has been cut off for greater clarity in the ladder limit, 
where the energetic separation of the singlet and triplet is now much 
smaller (Fig.~4). In the ladder regime one again sees singlet or triplet 
ground states due to the effective interaction between the free spins, 
whose sign is selected by the geometry. Here it is the triplet ground state 
which becomes a singlet as $\delta$ is reduced below 0.5, mirroring the 
behavior in Fig.~7(a,b). However, in this case both singlet ground states 
in the regime $0.25 < \delta < 0.5$ do not cross directly to the singlet 
states of the MG limit, but are separated from these by a region with very 
weak triplet ground states for both ``even'' and ``odd'' impurity 
configurations, at $0.15 < \delta < 0.23$. This triplet region is found to 
be robust in chains of 18 and 22 sites, which do not alter the short 
inter-impurity path, but is not present for the nearest accessible 
impurity configuration, which is two impurities on the first and third rungs. 

From above, the explanation of the triplet regime is expected within a 
free-spin scenario. In contrast to the ((1,b)(4,m)) case, the ground state 
for impurity configurations such as ((1,b)(5,m)) (Fig.~8), with odd numbers 
of dimers in the inter-impurity chain segment, should not be affected by 
the physics of dimer repolarization since the number of inter-dimer bonds 
is even. Indeed the dimerization patterns ${\cal S}_1$ and ${\cal S}_2$ 
are characteristic of the ladder polarization for dimerization values 
$\delta > 0.08$ for this length of segment. In the vicinity of the triplet 
regime, the patterns are qualitatively the same on both sides of the 
crossings at $\delta = 0.15$ and $\delta = 0.23$, and do not indicate a 
difference in the nature of the ground state. The extremely close competition 
between singlet and triplet reflected in the very small values of their 
separation in this regime suggests that the emergence of a triplet ground 
state is a higher-order effect which cannot be explained using only the bond 
spin correlation functions ${\cal S}_m$ or on-site expectation values 
$\langle S_z \rangle$. We have found a triplet ground state also for the 
open system of two ``impurity'' spins coupled to six spins in a frustrated 
and dimerized segment when $\delta = 0.2$. We have further verified using a 
10-spin intermediate segment that this triplet ground state is not present 
for longer inter-impurity segments with odd dimer number, although in this 
case the triplet energy approaches to within 2$\times$10$^{-6} J_1$ of 
the singlet. 

To summarize this section, we have analyzed the evolution of the free 
spins induced by nonmagnetic impurities in the spin ladder on tuning the 
dimerization to the MG limit. As frustration is increased, free-spin 
nature survives down to $\delta \sim 0.1$, below which the free spins 
are screened and the low-energy sector changes rapidly to reflect the 
dominance of local dimer formation and rearrangement. Although the free 
spins are only weakly bound in the ladder case, while all neighbouring 
spins are strongly bound in the MG chain, in fact the overall magnitude 
of the effective coupling is reduced towards zero between these limits by 
the presence of frustration, and its sign changes according to details of 
the impurity locations. 

\section{Phase Diagram}

For the thermodynamic limit of infinite chains with 
random doping by a finite concentration of nonmagnetic impurities, the 
system will contain finite segments of all possible inter-impurity spacings 
governed by a Poisson distribution.\cite{rmmb} Our results allow us to 
propose the schematic phase diagram in Fig.~10(a). For the majority of 
the parameter space, the system is dominated by geometrical dimer 
formation, and for all but the lowest energy scales will exhibit a free 
spin degree of freedom per impurity as these dimers are broken. The 
low-energy thermodynamic response would show the presence of these spins 
and the absence of a true spin gap. For small dimerization and appreciable 
frustration, there is a significant regime where the impurities are 
screened and do not feature in the thermodynamic response of the system, 
which as a result may still be appropriately characterized as a spin liquid. 
For a finite region of small $J_2$ and $\delta$, we anticipate an average 
of half a free spin per impurity, simply as the result of producing 
effectively isolated chain segments of odd and even lengths. Because no 
true transitions are possible in the system at finite doping, these 
regions should be connected by crossover regimes, where we cannot exclude 
the possibility of a mixed behavior, in that configuration-dependent local 
dimer formation may favor screening of some impurities while isolated spins 
are left around others. 

\begin{figure}[ht]
\centerline{\psfig{figure=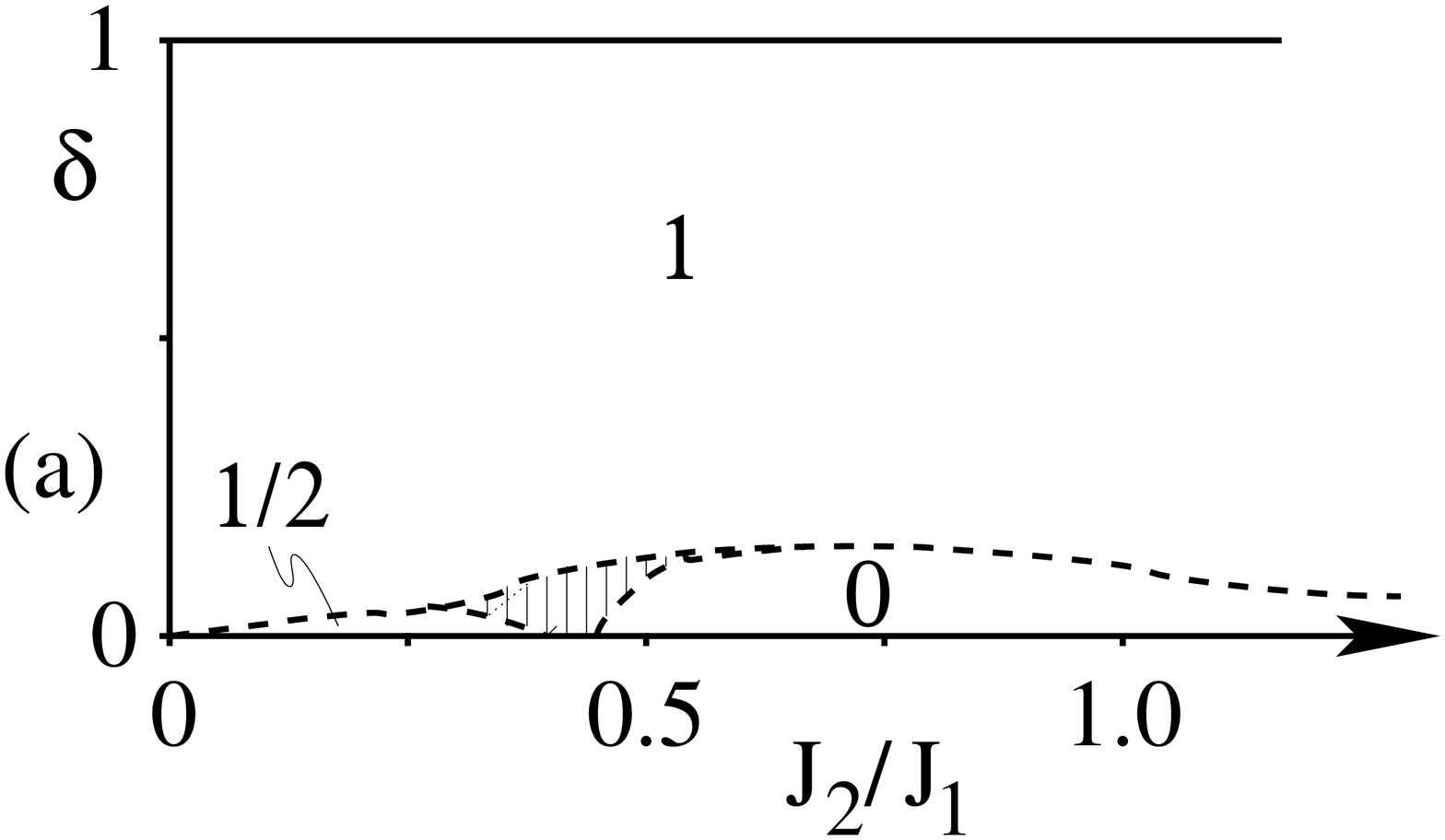,width=8.5cm,angle=0}}
\vspace{0.3cm}
\centerline{\psfig{figure=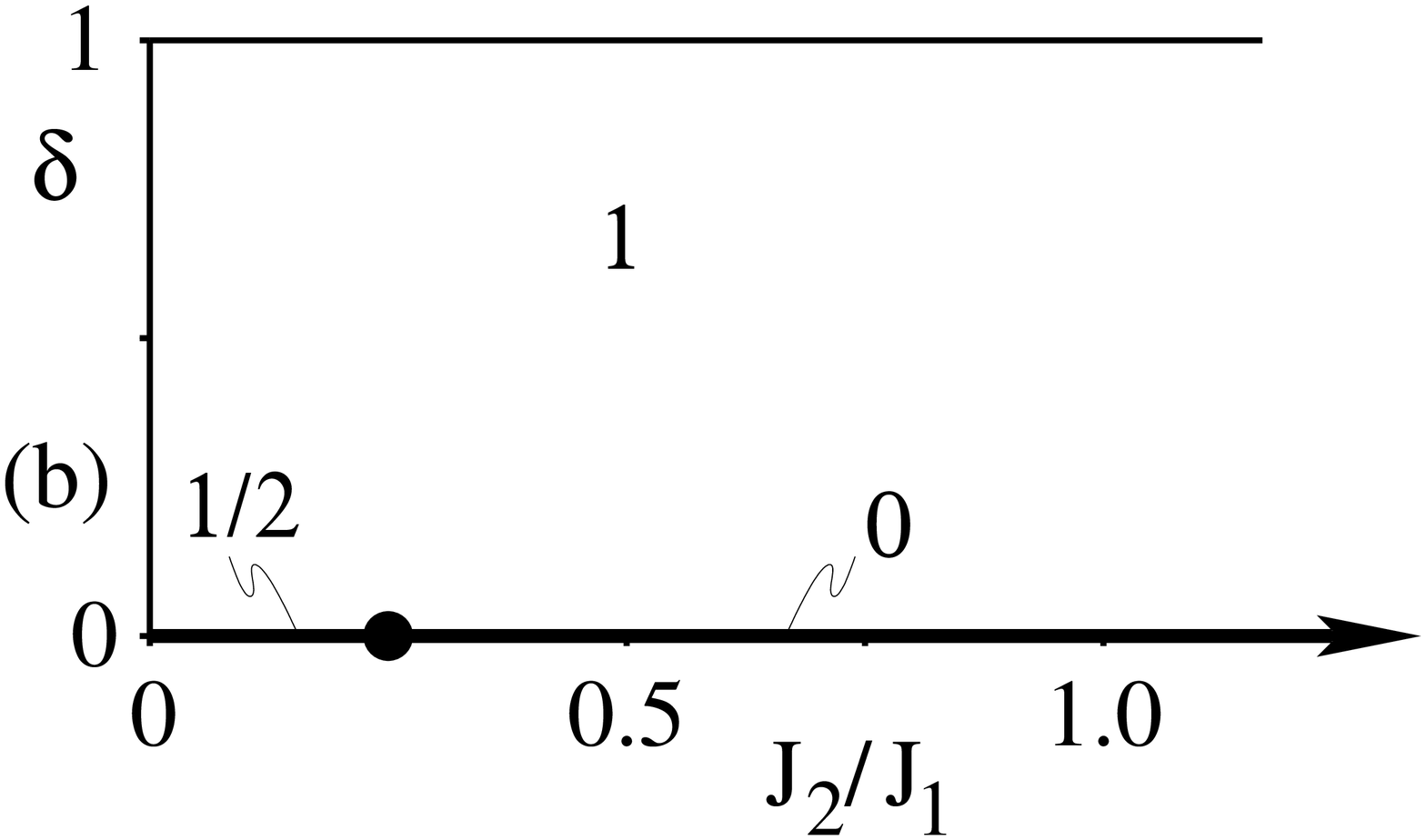,width=8.5cm,angle=0}}
\vspace{0.3cm}
\caption{ Schematic phase diagram showing the number of free spins per 
impurity for the dimerized, frustrated spin chain in the thermodynamic 
limit. (a) Finite impurity concentration: the dashed lines do not represent 
true phase transitions, but rather denote crossovers between different 
regimes of behavior. (b) Limit of vanishing impurity concentration. }
\end{figure}

Fig.~10(b) shows by contrast the phase diagram which we conjecture for a 
finite number of impurities, whose concentration vanishes in the thermodynamic 
limit. We stress that the results ${\cal S}_1 < 0$ and ${\cal S}_2 > 0$ 
which we obtain for the inter-dimer correlations in the MG chain must 
depend on the presence of impurities, as the exactness of the dimerized 
wave function guarantees that these quantities are zero in a pure chain. 
In the limit of infinitely separated impurities, any finite dimerization 
$\delta$ will then act to confine spin excitations, and place the system 
in the regime of free-spin physics. This statement may be justified by 
considering the persistence of an exact dimer wave function, without double 
degeneracy, along the line $2 J_2 / J_1 + \delta = 1$. Thus in the somewhat 
hypothetical situation of vanishing impurity concentration, the regions 
where an impurity introduces no free spins, or half of one, would be reduced 
to the $\delta = 0$ axis [Fig.~10(b)], and the three regions would be 
separated by true transitions.

\section{\mbox{\boldmath ${\rm C{\lowercase{u}}G{\lowercase{e}}O}_3$}}

One possible experimental realization of the dimerized, frustrated chain 
is the compound CuGeO$_3$, which is best known for exhibiting a spin-Peierls 
transition. In this material the dimerization is weak, and the frustration 
remains ambiguous due to competing models, both one- and two-dimensional, 
which have been used to fit the measured spin susceptibility. It is important 
to note also that in our model above, the dimerization parameter $\delta$ 
was taken to be rigid, meaning unaffected by the introduction of a 
nonmagnetic impurity, which corresponds to the limit of very high phonon 
frequencies. A further point of contention in CuGeO$_3$ is the rigidity of 
the lattice, or the extent to which the local dimerization pattern is in 
fact relaxed around impurity sites.\cite{rdhrap} 

The above uncertainties notwithstanding, the most direct experimental 
measure of the number of free spins in a system is the Curie-Weiss tail 
in the low-temperature susceptibility. This has been reported to scale 
with the impurity concentration in some ladder systems and 
in the spin-Peierls phase of CuGeO$_3$, implying the creation 
of one free spin per impurity. However, highly doped samples of CuGeO$_3$, 
which do not undergo a spin-Peierls transition, do not show a significant 
Curie-Weiss tail, and in any case nothing which may be scaled with the 
impurity concentration.\cite{rgrvpdr} We suggest that the presence of 
frustration, which is in addition likely to be stronger in the undimerized 
phase than in the spin-Peierls phase, is a viable candidate to explain this 
situation. 

Our analysis implies a lower bound on this frustration of $J_2 / J_1 > 
0.25$, a conclusion which cannot be drawn from the susceptibility alone. 
A further qualitative deduction is that frustrating interactions are much 
more significant than interchain interactions, which also act to stabilize 
the dimer pattern and free-spin physics, in this doping regime. Finally, 
we note that an impurity-screening mechanism due only to lattice relaxation 
would result on average in screening of only half of the impurities.

\section{Conclusion}

In summary, we have analyzed the behavior on doping by nonmagnetic 
impurities of the dimerized, frustrated Heisenberg spin chain. We 
find that when frustration is predominant it can act to screen the 
impurity-induced spins, even in systems with a spin gap. This is in 
contrast to the cases of the spin ladder and dimerized chain, where one 
impurity acts to introduce one free spin. In the general case of a 
chain with dimerization and frustration, free-spin physics around the  
impurities breaks down when the dimerization is reduced below $\delta 
\sim 0.1$, to be replaced by a screening of the impurities which arises 
from the near-degeneracy of the $J_1$ bonds. The sign and magnitude of 
the effective interactions show a complex but systematic evolution between 
the limiting cases. For long chains with finite impurity concentrations we 
deduce a phase diagram with crossovers, but no true phase transitions, 
separating regions where each dopant introduces 0, 1/2, or 1 free-spin 
degree of freedom. 

More generally, we may conclude that nonmagnetic impurities in quantum 
magnets give rise to a variety of subtle phenomena. The essential feature
of the undoped MG chain which leads to screening of the impurity sites is 
the presence of two degenerate singlet ground states with a finite gap to 
the first triplet. We may expect similar screening effects, leading to an 
absence of free-spin degrees of freedom, to operate around nonmagnetic 
impurities in other frustrated spin systems with analogous properties. A 
notable example in this category is the $S$ = 1/2 Heisenberg model on the 
Kagom\'e lattice, where the ground-state degeneracy takes the form of a 
continuum of low-lying singlets in the singlet-triplet gap.\cite{rlblps} 
Because a resonant valence-bond (RVB) formulation based on singlet dimer 
coverings has been shown to give a good description of these low-lying 
singlets,\cite{remmm} the same general screening mechanism, which is based 
on rearrangement of singlets around the impurities, is expected to apply 
also to this case.

\section*{Acknowledgments}

We are grateful to B. Grenier and A. P. Kampf for helpful discussions 
concerning CuGeO$_3$, and to M. Vojta for critical reading of the 
manuscript. We acknowledge financial support from SFB 484 of the Deutsche 
Forschungsgemeinschaft (BN) and from the Swiss National Fund (FM).


\begin{references}

\bibitem{rbhsl} N. Bulut, D. Hone, D. J. Scalapino, and E. Y. Loh, 
Phys. Rev. Lett. {\bf 62}, 2192 (1989).

\bibitem{rudjssvz} M. Ulmke, P. J. H. Denteneer, V. Janis, R. T. 
Scalettar, A. Singh, D. Vollhardt, and G. T. Zimanyi, Adv. Solid State 
Phys. {\bf 38}, 369 (1999).

\bibitem{rmacm} A. V. Mahajan, H. Alloul, G. Collin, and J.-F. Marucco, 
Phys. Rev. Lett. {\bf 72}, 3100 (1994).

\bibitem{rbmambcm} J. Bobroff, W. A. MacFarlane, H. Alloul, P. Mendels, 
N. Blanchard, G. Collin, J.-F. Marucco, Phys. Rev. Lett. {\bf 83}, 4381 
(1999).

\bibitem{rjfhbbscm} M.-H. Julien, T. Feher, M. Horvatic, C. Berthier, O. N. 
Bakharev, P. Segransan, G. Collin, and J.-F. Marucco, Phys. Rev. Lett. {\bf 
84}, 3422 (2000).

\bibitem{rvbs} M. Vojta, C. Buragohain, and S. Sachdev, Phys. Rev. B
{\bf 61}, 15152 (2000).

\bibitem{rhrg} S. Haas, J. Riera, and E. Dagotto, Phys. Rev. B {\bf 48}, 
13174 (1993).

\bibitem{rwfsl} E. Westerberg, A. Furusaki, M. Sigrist, and P. A. Lee,
Phys. Rev. Lett. {\bf 75}, 4302 (1995).

\bibitem{rsf} M. Sigrist and A. Furusaki, J. Phys. Soc.
Jpn. {\bf 65}, 2385 (1996).

\bibitem{rfts} H. Fukuyama, T. Tanimoto and M. Saito, J. Phys. Soc. 
Jpn. {\bf 65}, 1182 (1996). 

\bibitem{raftnt} M. Azuma, Y. Fujishiro, M. Takano, M. Nohara, and H. Takagi 
Phys. Rev. B {\bf 55}, R8658 (1997).

\bibitem{rskumhhs} Y. Sasago, N. Koide, K. Uchinokura, M. C. Martin, 
M. Hase, K. Hirota, and G. Shirane, Phys. Rev. B 54, R6835 (1996).

\bibitem{rdhrap} A. Dobry, P. Hansen, J. Riera, D. Augier, and D. 
Poilblanc, Phys. Rev. B {\bf 60}, 4065 (1999). 

\bibitem{rwnsh} S. Wessel, B. Normand, M. Sigrist, and S. Haas, 
Phys. Rev. Lett. {\bf 86}, 1068 (2001).

\bibitem{rmlrd} G. B. Martins, M. Laukamp, J. Riera, and E. Dagotto, 
Phys. Rev. Lett. {\bf 78}, 3563 (1997).

\bibitem{rlmgmdhlr} M. Laukamp, G. B. Martins, C. Gazza, A. L. Malvezzi, 
E. Dagotto, P. M. Hansen, A. C. Lopez, and J. Riera, Phys. Rev. B {\bf 57}, 
10755 (1998).

\bibitem{rm} W. Marshall, Proc. R. Soc. London Ser. A, {\bf 232}, 48 (1955).

\bibitem{rmmb} F. Mila, P. Millet, and J. Bonvoisin, Phys. Rev. B {\bf 54}, 
11925 (1996).

\bibitem{rgrvpdr} B. Grenier, J.-P. Renard, P. Veillet, C. Paulsen, G. 
Dhalenne, and A. Revcolevschi, Phys. Rev. B {\bf 58}, 8202 (1998); B. 
Grenier, J.-P. Renard, P. Veillet, L.-P. Regnault, J. E. Lorenzo, C. 
Paulsen, G. Dhalenne, and A. Revcolevschi, Physica B {\bf 259-261}, 954 
(1999). 

\bibitem{rlblps} P. Lecheminant, B. Bernu, C. Lhuillier, L. Pierre, and 
P. Sindzingre, Phys. Rev. B {\bf 56}, 2521 (1997). 

\bibitem{remmm} V. Elser, Phys. Rev. Lett. {\bf 62}, 2405 (1989); 
F. Mila, Phys. Rev. Lett. {\bf 81}, 2356 (1998); M. Mambrini and F. Mila, 
Eur. Phys. J. B {\bf 17}, 651 (2000).

\end{references}
\end{document}